\documentclass[reprint, amsmath,amssymb, aps, longbibliography, pre]{revtex4-2}
\usepackage[T1]{fontenc}
\usepackage[utf8]{inputenc} 
\usepackage{appendix}
\usepackage{graphicx}
\usepackage{dcolumn}
\usepackage{bm}
\usepackage{hyperref}
\usepackage{csquotes}
\usepackage{booktabs}
\usepackage{xcolor}
\usepackage{comment}
\usepackage[normalem]{ulem}

\newcommand{\rsn}{r_\mathrm{SN}}
\newcommand{\roc}{r_{0\mathrm{c}}}
\newcommand{\ric}{r_{1\mathrm{c}}}
\newcommand{\rhosn}{\rho_{\mathrm{SN}}}
\newcommand{\keepmath}[1]{\mbox{\ensuremath{#1}}}

\begin{document}

\preprint{APS/123-QED}

\title{Emergence of cooperation in nonlinear higher-order public goods games}

\author{Jaume Llabrés}
\email{jaumellabres@ifisc.uib-csic.es}
\affiliation{Institute for Cross-disciplinary Physics and Complex Systems IFISC (CSIC-UIB), Campus Universitat Illes Balears, 07122 Palma de Mallorca, Spain.}
\affiliation{Department of Network and Data Science, Central European University Vienna, Vienna 1100, Austria}

\author{Onkar Sadekar}
\affiliation{Department of Network and Data Science, Central European University Vienna, Vienna 1100, Austria}
\affiliation{Human Evolutionary Ecology Group, Department of Evolutionary Anthropology, University of Zurich, Zurich, Switzerland}

\author{Federico Malizia}
\affiliation{Department of Network and Data Science, Central European University Vienna, Vienna 1100, Austria}

\author{Federico Battiston}
\email{battistonf@ceu.edu}
\affiliation{Center for Networks and Complexity, Luiss University of Rome, Rome, Italy}
\affiliation{Department of AI, Data and Decision Sciences, Luiss University of Rome, Rome, Italy}
\affiliation{Department of Network and Data Science, Central European University Vienna, Vienna 1100, Austria}

\date{\today}

\begin{abstract}
Evolutionary game theory has provided substantial contributions to explain the emergence of cooperation under unfavorable conditions in ecology, economics, and the social sciences.
Recently, inspired by newly available empirical evidence on group interactions, higher-order networks have emerged as a natural framework to encode multiplayer games in structured populations.
Here, we study the emergence of cooperation in nonlinear public goods games on hypergraphs, where collective reinforcement captures the synergistic or discounting effect associated with each additional cooperator. 
In well-mixed populations, when all games have the same number of players, the system displays a transition whose nature changes from continuous to discontinuous depending on the form of nonlinearity. By contrast, when games involving different numbers of players coexist, an additional bistability region between an active coexistence state and full cooperation may emerge.
We further find that scale-free hypergraphs promote cooperation, highlighting the crucial role played by both the initial placement of cooperators and the presence of hyperdegree correlations.
Overall, our results provide a comprehensive characterization of nonlinear public goods games on hypergraphs and open avenues for richer models of evolutionary dynamics of multiplayer games on structured populations.
\end{abstract}

\maketitle

\section{Introduction}

Cooperation in large groups of unrelated individuals is a hallmark of human evolution~\cite{ nowak_supercooperators_2012, christakis_friendship_2014, christakis2019blueprint}. The emergence and stability of such pro-social behavior is an active area of research in diverse fields such as economics, biology, physics, and psychology~\cite{von1944theory, smith_logic_1973, hauert2005game, perc2017statistical}. Game theory provides a robust mathematical backbone to study the strategic behavior of players by assigning \textit{payoffs} for each possible combination of player actions~\cite{von1944theory, nash1950equilibrium}. Based on the relative magnitudes of possible payoffs, collective action problems emerge naturally. Such problems depict a scenario where the best action for an individual (in terms of their payoff) does not align with the best action for the group~\cite{hardin_tragedy_1968, maynard_smith_evolution_1982, dawkins_selfish_2006}. 

Social dilemmas are a type of collective action problem where each individual chooses between two strategies -- cooperate and defect. Cooperation provides benefits to the group at a personal cost, while defection allows for free-riding on other cooperators' contributions. The prisoner's dilemma and its multiplayer equivalent -- the public goods game (PGG) -- are paradigmatic examples of such collective action problems. However, players repeatedly interact with other players and can change their strategies based on the relative benefits of various competing strategies. Building on the idea of evolution by natural selection where only the \textit{fittest survive}, evolutionary game theory provides a natural way to study the spread and evolution of cooperative behavior based on the fitness of strategies~\cite{taylor_evolutionary_1978, hofbauer_evolutionary_1998}. Under such a framework, strategies which provide a higher fitness value to individuals (in terms of payoffs) are preferentially adopted in the population and can lead to the survival of cooperation.

A central question is how the outcome of these dilemmas is shaped by the structure of interactions. While well-mixed populations provide a robust framework to uncover insights into the collective behavior of individual agents, they do not truly reflect the real-life organization of social interactions ~\cite{albert2002statistical, newman2018networks}. These interaction patterns are commonly represented by graphs when interactions are pairwise~\cite{boccaletti2006complex}, and by hypergraphs when they involve groups of individuals~\cite{berge1984hypergraphs, battiston2020networks, battiston2021physics, battiston2025higher, battiston2026collective}. These structured patterns can strongly affect both the emergence of cooperation and the resulting collective dynamics, often giving rise to nontrivial forms of self-organization and phase transitions~\cite{pastor2001epidemic, santos2005scale, santos2006evolutionary, nowak1992evolutionary, szabo2007evolutionary, nowak2006five}. 
For instance, depending on the specific rule for evolution of strategies, network reciprocity facilitates cooperative agents to invade and fixate on graphs~\cite{lieberman2005evolutionary, ohtsuki2006simple, allen2017evolutionary}. Furthermore, specific network features, such as scale-free degree distributions, can promote cooperation in adverse settings based on the influence of \textit{hubs}~\cite{santos2008social, perc2013evolutionary}. 

A major advantage of considering non-pairwise interactions is that they provide a natural microscopic framework to account for synergistic or discounting group benefits~\cite{Hauert2006}. Even though it is possible to model multiplayer games on graphs by considering a node and its neighbors to be a part of a group~\cite{santos2008social}, many interesting structural aspects such as overlapping interactions are not straightforward to consider. As such, graph-based structures are not the best way to model multiplayer games. Hypergraphs offer an intuitive way to model multiplayer games by considering each hyperedge as a separate game~\cite{alvarez2021evolutionary, civilini2024explosive, battiston2025higher}. Recent works have shown that multiplayer games on hypergraphs allow us to explore new questions and uncover new mechanisms to understand the emergence of cooperation~\cite{alvarez2021evolutionary, burgio2020evolution, guo2021evolutionary, gao2025evolutionary}. In particular, Civilini \textit{et al.} introduced a general framework for two- and three-player games on hypergraphs and found that increasing the fraction of three-player games paved the way for an abrupt transition to a bistable majority cooperation state accompanied by a full defection state \cite{civilini2024explosive}. 
On the other hand, Sheng \textit{et al.} considered a nonlinear PGG generalizable to any number of players on arbitrary hypergraphs~\cite{Sheng2024}. They found that higher-order networks promote cooperation more than well-mixed populations and pairwise networks by calculating the fixation probability under weak selection and death-birth updating rule. 
A nonlinear PGG with heterogeneous synergistic and discounting interactions has been studied~\cite{zhu2024}, reporting the emergence of coexistence between active and cooperative states as well as bistable regimes. However, the interplay between heterogeneous nonlinear responses and interactions of different orders remains unexplored.
Subsequent works have further explored the drivers of cooperation in group structured populations~\cite{sadekar2025drivers, guo2025evolutionary, wang2025emergence, wang2026evolution, wang2026strategy}.

Nonlinearity plays a central role in modeling social dynamics and can generate rich dynamical effects, inducing bistability and abrupt transitions in various processes~\cite{iacopini2019simplicial, skardal2019abrupt, neuhauser2020multibody, fan2022coupled, robiglio2025higher, perez2025social, zhang2023impact}. For evolutionary games, nonlinearity arises naturally through the payoffs of multiplayer interactions, where each additional cooperator nonlinearly affects the total group benefit. This generalization was first studied to understand the effects of synergy and discount, recovering all commonly studied social dilemmas in well-mixed~\cite{Hauert2006} and networked populations~\cite{Li2015}. Despite these advances, the interplay between nonlinear multiplayer interactions and population structure is still not fully understood. Most previous studies have focused either on pairwise games on networks or on single-order multiplayer games. However, real social interactions typically involve groups of different sizes, so that strategic interactions of different orders coexist within the same population. How the competition between different interaction orders affects the evolutionary dynamics of cooperation is still largely unexplored.

Also, much less is known about how the structural organization of hypergraphs influences the outcome of evolutionary game dynamics. While these effects have been extensively explored in other higher-order dynamical processes~\cite{landry2020effect,st2022influential, zhang2023higher, malizia2025disentangling, lucas2026reducibility}, they remain far less understood in the context of evolutionary games.

To address these questions, in this work we study a nonlinear PGG with a mixture of two orders of interaction on hypergraphs. We first derive an analytical mean-field description of the evolutionary dynamics showing that the coexistence between two different orders can generate a qualitatively richer phase structure than in single-order games. We then investigate how these phenomena are modified in structured populations represented as hypergraphs. The outline of the paper is as follows. In Sec.~\ref{sec:model_pgg} we introduce the model. In Sec.~\ref{sec:evo_dyn} we analyze the evolutionary dynamics and determine the stationary states for both single-order and mixed-order games in well-mixed populations. In Sec.~\ref{sec:hypergraph} we study the dynamics on hypergraphs and quantify the deviations from the well-mixed scenario. Finally, in Sec.~\ref{sec:discussion}, we discuss the results and outline possible directions for future research.

\section{The model: Nonlinear Public Goods Games}
\label{sec:model_pgg}
We consider a population of $N$ players represented as a hypergraph $\mathcal{H}(\mathcal{V},\mathcal{E})$, where $\mathcal{V}$ denotes the set of nodes and $\mathcal{E}$ the set of hyperedges, where a hyperedge of size $\ell$, or $\ell$-hyperedge, connects a subset of $\ell$ different nodes. Each player $i \in \mathcal{V}$ is endowed with a binary strategy variable $s_i \in \{0,1\}$, where $s_i=1$ represents cooperation (C) and $s_i=0$ defection (D). Interactions take place in groups defined by hyperedges of size $\ell\geq 2$, each representing a game involving $\ell$ players, or an $\ell$-player game. In every such game, cooperators contribute an amount $c>0$ to a common pool, whereas defectors contribute nothing. The total benefit produced by the group is then equally shared among all participants~\cite{marwell1979experiments}. 

For an $\ell$-player game containing $n$ cooperators, the collective benefit is assumed to depend nonlinearly on $n$ according to~\cite{Hauert2006}
\begin{equation}
    b_n^{(\ell)} 
    = b \left( 1 + \delta_\ell + \delta_\ell^2 + \cdots + \delta_\ell^{\,n-1} \right)
    = b\,\frac{1-\delta_\ell^{\,n}}{1-\delta_\ell},
\end{equation}
where $b>0$ sets the overall benefit scale and $\delta_\ell$ controls the degree of nonlinearity of the interaction. For $\delta_\ell<1$, the game exhibits discounting interactions with diminishing marginal returns; $\delta_\ell=1$ corresponds to the linear PGG with $b_n^{(\ell)}=b\,n$; and $\delta_\ell>1$ describes synergistic interactions with increasing marginal returns. Accordingly, the payoffs associated with an $\ell$-player game with $n$ cooperators are given by
\begin{subequations}
\begin{align}
    \pi^{(\ell)}_{\mathrm{C}} &= \frac{b_n^{(\ell)}}{\ell} - c,\\
    \pi^{(\ell)}_{\mathrm{D}} &= \frac{b_n^{(\ell)}}{\ell}.
\end{align}
\end{subequations}

The evolutionary dynamics follows a pairwise comparison update rule \cite{traulsen2006stochastic, santos2006evolutionary, perc2013evolutionary}. At each time step, a focal player $f$ with strategy $s_f$ is selected uniformly at random. Then, a model player $m$, with strategy $s_m$, is selected uniformly at random from the other members of a randomly chosen hyperedge to which $f$ belongs. Both players accumulate their total payoff by participating in all games associated with the hyperedges to which they belong. Let $\pi_f$ and $\pi_m$ denote their total payoffs. The focal player adopts the model's strategy with probability
\begin{equation}
    p(s_f\to s_m)=\frac{1}{1+e^{-w(\pi_m-\pi_f)}},
\end{equation}
where $w$ is the strength of the selection.
\begin{figure}
    \centering
\includegraphics[width=0.49\textwidth]{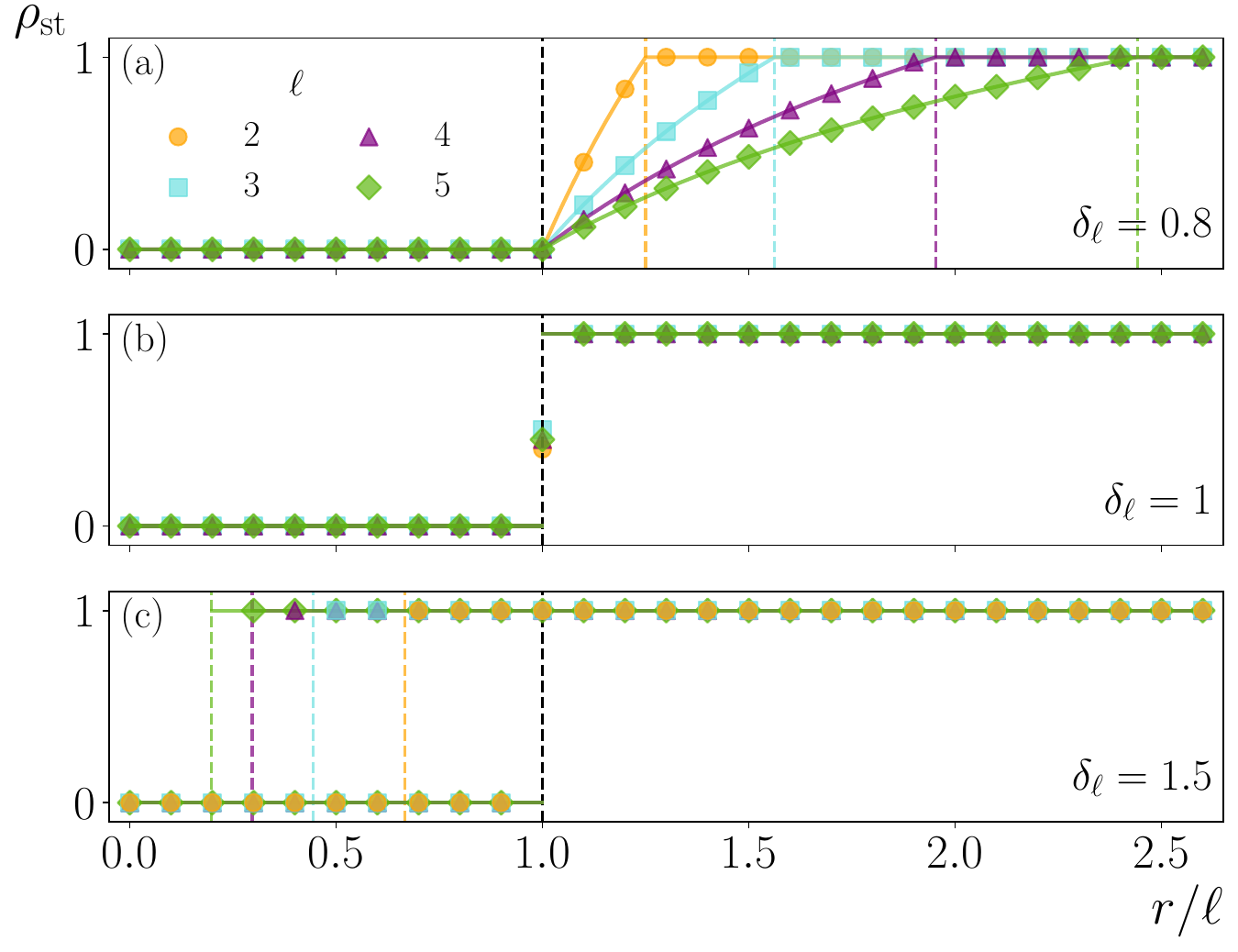}
    \caption{\textbf{Single-order $\ell$-player nonlinear public goods game in well-mixed populations.} Stationary density of cooperators $\rho_{\mathrm{st}}$ versus the rescaled benefit-to-cost ratio $r/\ell$ for several group sizes $\ell$ and the three different regimes $\delta_\ell<1$, $\delta_\ell=1$, and $\delta_\ell>1$, as indicated. For the linear case $\delta_\ell=1$, all curves collapse. Symbols correspond to results obtained from computer simulations in the well-mixed population for $N=~12000$ while solid lines correspond to the analytical solutions of the mean-field replicator equation, Eq.~\eqref{eq:drhodt_l}. Black dashed lines indicate the stability threshold $\roc^{(\ell)}/\ell=1$, while colored dashed lines correspond to $\ric^{(\ell)}/\ell=\delta_\ell^{-(\ell-1)}$ for each group size $\ell$, respectively.}
\label{fig:PD_l}
\end{figure}

\begin{figure*}[ht!]
    \centering
    \includegraphics[width=\textwidth]{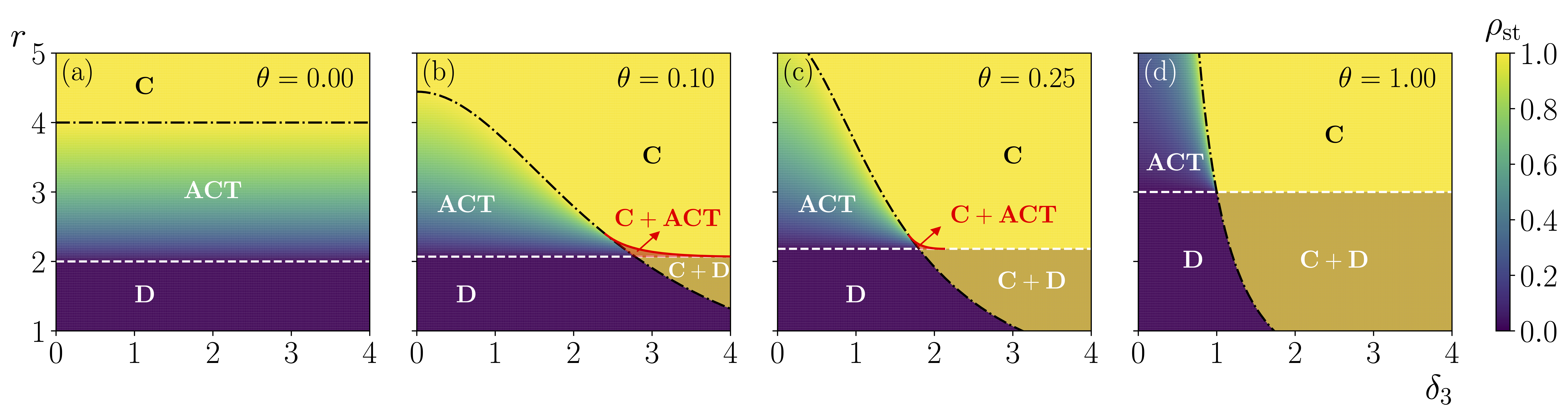}
    \caption{\textbf{Mixed-order two- and three-player nonlinear public goods games in well-mixed populations.} Phase diagrams of the stationary density of cooperators $\rho_\mathrm{st}$ in the $(\delta_3,r)$ plane for fixed pairwise nonlinearity $\delta_2=0.5$ and several values of the fraction of three-player games $\theta$, as indicated in each panel. White and black dashed lines correspond to the stability thresholds $\roc(\theta)$ and $\ric(\theta,\delta_2,\delta_3)$, defined in Eqs.~\eqref{eq:r0c_r1c_l=2,3}, respectively. The red line denotes the saddle-node line $\rsn(\theta,\delta_2,\delta_3)$, which appears when the two interior fixed points collide at $\rho=\rhosn$, given by Eq.~\eqref{eq:rho_sn}, within the physical interval $[0,1]$. The different regions correspond to stable solutions of full defection (D), full cooperation (C), active coexistence (ACT), bistability between full cooperation and full defection (C+D), and bistability between full cooperation and an interior active state (\mbox{C+ACT}).
    }
    \label{fig:PD_mixed_l2_l3}
\end{figure*}

\section{Mean-field evolutionary dynamics} \label{sec:evo_dyn}
In this section, we analyze the stationary states of nonlinear PGGs in well-mixed populations by deriving the corresponding mean-field replicator equation. We first review the single-order case, in which all interactions correspond to $\ell$-player games.
We then extend the analysis to mixed-order nonlinear PGGs combining $\ell_1$- and $\ell_2$-player games, before focusing on the particular case of pairwise $(\ell_1=2)$ and triplet $(\ell_2=3)$ games.

\subsection{Single-order $\ell$-player nonlinear public goods games}
\label{subsec:single_l}

Let $\rho\in[0,1]$ denote the fraction of cooperators in the population.
In the mean-field limit, its time evolution is governed by the replicator equation \cite{hofbauer_evolutionary_1998}
\begin{equation}
    \label{eq:drhodt_l_1}
    \frac{d\rho}{dt} = \rho(1-\rho)\,\Delta\pi^{(\ell)},
\end{equation}
where $\Delta\pi^{(\ell)}\equiv\pi^{(\ell)}_{\mathrm C}-\pi^{(\ell)}_{\mathrm D}$ is the payoff difference between cooperators and defectors when playing an $\ell$-player game. The expected payoffs of cooperators and defectors are, respectively, given by
\begin{subequations}
\label{eq:pis_l}
\begin{align}
    \pi^{(\ell)}_{\mathrm C}
    &= -c + \frac{b}{\ell(1-\delta_\ell)}
       \sum_{k=0}^{\ell-1} \mathrm{B}(k, \ell-1;\rho)
       \left(1-\delta_\ell^{\,k+1}\right), \\
    \pi^{(\ell)}_{\mathrm D}
    &= \frac{b}{\ell(1-\delta_\ell)}
       \sum_{k=0}^{\ell-1} \mathrm{B}(k, \ell-1;\rho)
       \left(1-\delta_\ell^{\,k}\right),
\end{align}
\end{subequations}
where $\mathrm{B}(k, \ell-1;\rho)$ is the probability that a focal individual interacts with $k$ cooperators among the remaining $\ell-1$ members of the group. In well-mixed populations, this probability follows a binomial distribution,
\begin{equation}
    \label{eq:Binomial}
    \mathrm{B}(k, \ell-1;\rho)
    = \binom{\ell-1}{k}\rho^k(1-\rho)^{\ell-1-k},
\end{equation} 
which allows us to express the payoff difference as

\begin{equation} \label{eq:payoff_diff}
    \Delta\pi^{(\ell)}=
        \frac{b}{\ell}
        \bigl(1+(\delta_\ell-1)\rho\bigr)^{\ell-1} - c.
\end{equation}
Substituting Eq.~\eqref{eq:payoff_diff} into
Eq.~\eqref{eq:drhodt_l_1} yields
\begin{equation}
\label{eq:drhodt_l}
    \frac{d\rho}{dt}
    = \rho(1-\rho)
      \bigg[
        \frac{r}{\ell}
        \bigl(1+(\delta_\ell-1)\rho\bigr)^{\ell-1}
        - 1
      \bigg],
\end{equation}
where we have defined the benefit-to-cost ratio $r\equiv b/c$ and rescaled time as $t\to c\,t$. This dynamical system presents two absorbing fixed points $\rho=0$ and $\rho=1$, corresponding to full defection and full cooperation, respectively. The threshold values $r_{[0\mathrm{c},1\mathrm{c}]}^{(\ell)}$ at which these solutions change stability can be determined by means of a linear stability analysis as
\begin{align}
    \roc^{(\ell)} &= \ell,\label{eq:r0c_single}\\
    \ric^{(\ell)} &= \frac{\ell}{\delta_\ell^{\,\ell-1}},\label{eq:r1c_single}
\end{align}
for $\rho=0$ and $\rho=1$ respectively. Additionally, this dynamical system may admit an interior solution given by~\cite{Hauert2006}
\begin{equation}
\label{eq:rho_star_general}
    \rho^*
    = \frac{1}{\delta_\ell - 1}
      \left[
        \left(\frac{\ell}{r}\right)^{1/(\ell-1)} - 1
      \right].
\end{equation}

The nature of the phase transitions exhibited by the system depends on the nonlinearity parameter $\delta_\ell$.
\begin{itemize}
    \item \textbf{Sublinear case $(\delta_\ell<1)$:} The system exhibits a continuous transition from full defection ($r<\roc^{(\ell)}$) to full cooperation ($r>\ric^{(\ell)}$), given that $\roc^{(\ell)}<\ric^{(\ell)}$. For \keepmath{\roc^{(\ell)}<r<\ric^{(\ell)}}, there exists a stable interior fixed point, a scenario in which cooperators and defectors coexist and which corresponds to the generalization of the snowdrift game to $\ell$ players.
    \item \textbf{Linear case $(\delta_\ell=1)$:}
    $\roc^{(\ell)}=\ric^{(\ell)}=\ell$. No interior fixed point exists for $r\neq\ell$, whereas at $r=\ell$ every value of $\rho$ is stationary. The system undergoes a discontinuous phase transition at $r=\ell$ from full defection ($r<\ell$) to full cooperation ($r>\ell$).

    \item \textbf{Superlinear case $(\delta_\ell>1)$:} The system exhibits a discontinuous transition from full defection ($r<\ric^{(\ell)}$) to full cooperation ($r>\roc^{(\ell)}$), given that $\roc^{(\ell)}>\ric^{(\ell)}$. An unstable interior fixed point exists for $\ric^{(\ell)}<r<\roc^{(\ell)}$, yielding a bistability region of width $\Delta r = \ell - \ell/\delta_\ell^{\,\ell-1}$ between full defection and full cooperation. This corresponds to the generalization of the stag-hunt game to $\ell$ players.
\end{itemize}

The three regimes described above are illustrated in Fig.~\ref{fig:PD_l}, which shows the stationary density of cooperators $\rho_{\mathrm{st}}$ as a function of the rescaled benefit-to-cost ratio $r/\ell$ for different values of $\ell$. We find excellent agreement between stochastic simulations in well-mixed populations and analytical results obtained from Eq.~\eqref{eq:drhodt_l}.

\subsection{Mixed-order $\ell_1$- and $\ell_2$-player nonlinear public goods games}

We now consider a population in which interactions of two different orders $\ell_1,\ell_2$ coexist. Specifically, individuals participate in an $\ell_1$-player game with probability $1-\theta$ and in an $\ell_2$-player game with probability $\theta$, where the corresponding nonlinearities are denoted by $\delta_{\ell_1}$ and $\delta_{\ell_2}$, respectively.

In this mixed-order setting, the mean-field replicator equation is obtained by considering the weighted contribution of the payoff differences associated with each interaction order,
\begin{equation}
\frac{d\rho}{dt}
=
\rho(1-\rho)
\left[
(1-\theta)\Delta\pi^{(\ell_1)}
+
\theta \Delta\pi^{(\ell_2)}
\right],
\label{eq:mixed_general}
\end{equation}
where the payoff difference $\Delta\pi^{(\ell)}$ is given by Eq.~\eqref{eq:payoff_diff}.

The central question is whether the coexistence of different interaction orders can generate stationary solutions that are absent in single-order $\ell$-player nonlinear PGGs. In particular, we would like to determine how many interior fixed
points can exist and under which conditions they emerge. To this end, we define
\begin{equation}
\label{eq:F_l1_l2}
    F(\rho)
    =
    (1-\theta)\Delta\pi^{(\ell_1)}
    +
    \theta \Delta\pi^{(\ell_2)},
\end{equation}
such that the interior fixed points are determined by the condition $F(\rho)=0$. The number and stability of the interior fixed points are therefore completely determined by the shape of $F(\rho)$ in the physical
interval $\rho\in[0,1]$.

As shown in Appendix~\ref{app:mixed_order}, $F(\rho)$ can have at most two isolated zeros in $(0,1)$. Moreover, two qualitatively different situations can be distinguished depending on the values of the nonlinearity parameters. When both interaction orders belong to the same regime, i.e., both are discounting ($\delta_{\ell_1},\delta_{\ell_2}<1$) or both are synergistic ($\delta_{\ell_1},\delta_{\ell_2}>1$), the function $F(\rho)$ remains monotonic and at most one interior fixed point can exist, recovering the same qualitative behavior as in single-order $\ell$-player nonlinear PGGs. In this case, mixing interaction orders only shifts the location of the transition lines and does not generate new dynamical regimes.

By contrast, when the two interaction orders belong to opposite regimes (\keepmath{\delta_{\ell_1}<1<\delta_{\ell_2}} or \keepmath{\delta_{\ell_2}<1<\delta_{\ell_1}}), $F(\rho)$ may become non-monotonic, and a saddle-node bifurcation may generate two interior solutions, one stable and one unstable. This leads to a bistable regime that is not present in single-order games.

In the following, we focus on the case $\ell_1=2$ and $\ell_2=3$, which allows for an analytically tractable study of the stationary solutions. For this case, the mean-field replicator equation takes the form
\begin{equation}
\label{eq:drhodt_mixed_23}
    \frac{d\rho}{dt}
    = \rho(1-\rho)
    \left[
        (1-\theta)\,\Delta\pi^{(2)}
      + \theta\,\Delta\pi^{(3)}
    \right],
\end{equation}
with
\begin{subequations}
\begin{align}
    \Delta\pi^{(2)}
    &= \frac{r}{2}\bigl[1+(\delta_2-1)\rho\bigr] - 1,
    \\
    \Delta\pi^{(3)}
    &= \frac{r}{3}\bigl[1+(\delta_3-1)\rho\bigr]^2 - 1.
\end{align}
\end{subequations}

The stability of the absorbing solutions $\rho=0$ and $\rho=~1$ is determined in this case by the critical values:
\begin{subequations}
    \label{eq:r0c_r1c_l=2,3}
    \begin{align} 
        \roc(\theta) & = \frac{6}{3-\theta}, \label{eq:r0c_l=2,3} \\
        \ric(\theta,\delta_2, \delta_3) &= \frac{6}{3\delta_2(1-\theta)+2\delta_3^2\theta}, \label{eq:r1c_l=2,3}
    \end{align}
\end{subequations}
such that for $\theta=0$ and $\theta=1$ we recover the critical values, given in Eqs.~\eqref{eq:r0c_single} and~\eqref{eq:r1c_single}, for the single-order two- and three-player nonlinear PGGs, respectively.

\begin{figure}[t]
    \centering
    \includegraphics[width=0.49\textwidth]{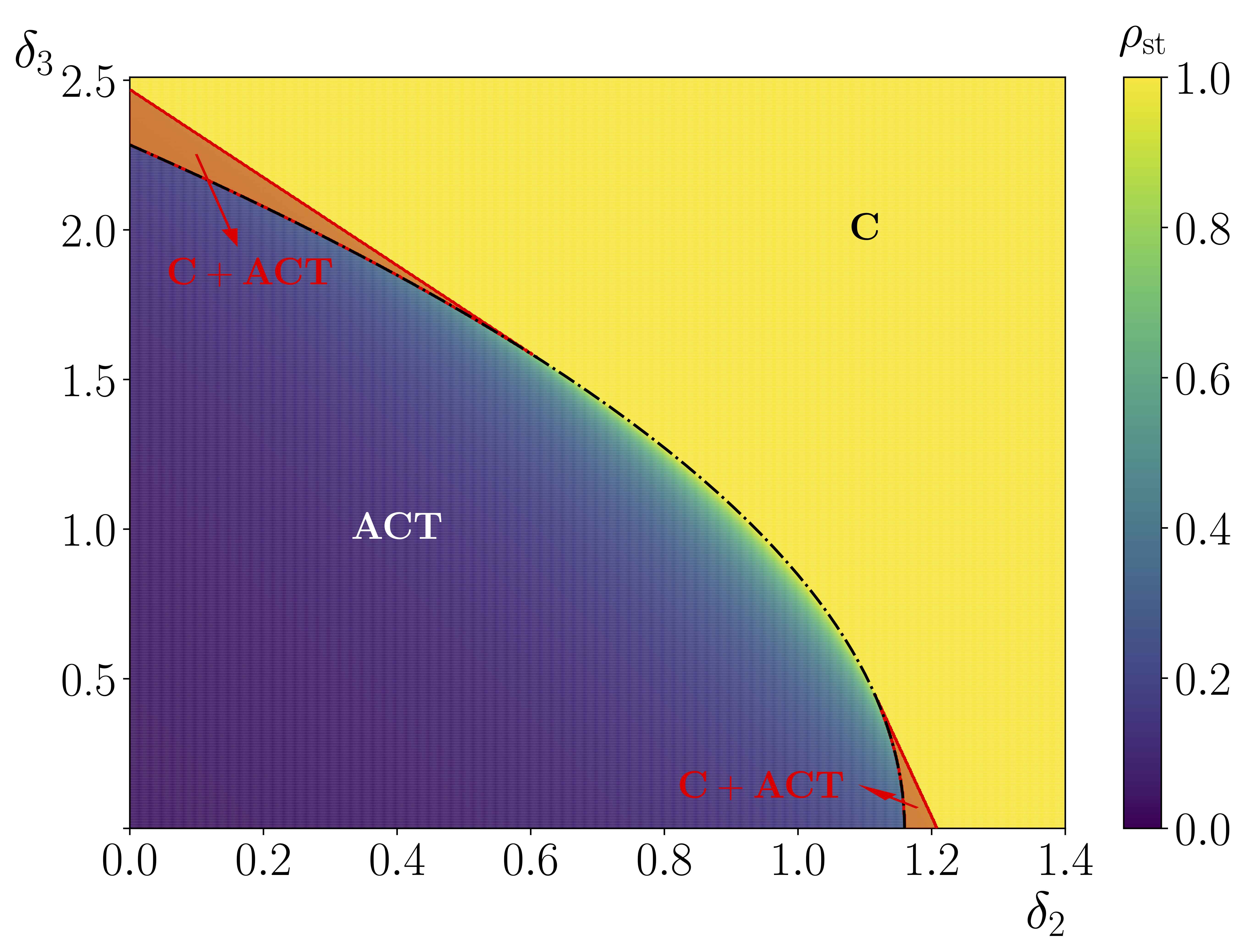}
    \caption{\textbf{Role of nonlinearities in the stationary cooperation state.} Phase diagram of the stationary density of cooperators $\rho_\mathrm{st}$ in the $(\delta_2,\delta_3)$ plane for $r=2.3$ and $\theta=~0.25$. Black dashed line corresponds to the stability threshold $\ric(\theta,\delta_2,\delta_3)$, defined in Eq.~\eqref{eq:r1c_l=2,3}. The red line denotes the saddle-node line $\rsn(\theta,\delta_2,\delta_3)$, which appears when the two interior fixed points collide at $\rho=\rhosn$, given by Eq.~\eqref{eq:rho_sn}, within the physical interval $[0,1]$. The different regions correspond to stable solutions of full cooperation (C), active coexistence (ACT), and bistability between full cooperation and an interior active state (\mbox{C+ACT}).}
    \label{fig:PD_delta2_delta3}
\end{figure}

As discussed above, the system may admit up to two interior fixed points, which in this particular case are given by
\begin{equation}
    \rho^*_\pm
    =
    \frac{
        r \bigl(3\delta_2(\theta-1) - 4\delta_3\theta + \theta + 3\bigr)
        \pm \sqrt{\Delta(\delta_2,\delta_3,\theta)}
    }{
        4 (\delta_3-1)^2\, \theta\, r
    },
\end{equation}
where the discriminant reads
\begin{align}
   \Delta(\delta_2,\delta_3,\theta)
   &= r\Bigl(
        r [3 \delta_2 (\theta -1)-4 \delta_3 \theta +\theta +3]^2
        \notag\\
        &\qquad\qquad
        +\, 8 (\delta_3-1)^2 \theta \bigl[(\theta -3) r+6\bigr]
   \Bigr).
\end{align}

The two branches $\rho^*_\pm$ exist only as long as \keepmath{\Delta(\delta_2,\delta_3,\theta) \ge 0}, which defines an upper threshold $\rsn(\theta,\delta_2,\delta_3)$ through the condition $\Delta(\delta_2,\delta_3,\theta)=0$. At this point, the two interior fixed points collide in a saddle-node bifurcation at
\begin{equation}
\label{eq:rho_sn}
    \rhosn
    = \frac{
        3\delta_2(\theta-1) + \theta(1-4\delta_3) + 3
    }{
        4 (1-\delta_3)^2 \theta
    }.
\end{equation}

In agreement with the general analysis presented above, if $\delta_2$ and $\delta_3$ share the same character, i.e., both sublinear or both superlinear, the saddle-node point lies outside the physical interval, $\rhosn\notin[0,1]$. The stationary behavior is therefore fully determined by the relative ordering of the stability thresholds $\roc(\theta)$ and $\ric(\theta,\delta_2,\delta_3)$. In this regime, the system qualitatively reproduces the phenomenology of the single-order $\ell$-player nonlinear PGG, with only a shift of the critical points.

When $\delta_2<1<\delta_3$ (or $\delta_3<1<\delta_2$), the saddle-node may enter the physical domain, $\rhosn\in[0,1]$, leading to a transition scenario that combines continuous and discontinuous features.
If this occurs, the system undergoes a continuous transition at $r=\roc(\theta)$, leading to the stable interior branch $\rho^*_-$. If $\roc(\theta)<\ric(\theta,\delta_2,\delta_3)$, this solution coexists with full cooperation in a region of bistability before terminating at \keepmath{r=\rsn(\theta,\delta_2,\delta_3)} through a saddle-node collision, which induces a discontinuous transition to $\rho=1$. If instead \keepmath{\ric(\theta, \delta_2, \delta_3)<\roc(\theta)}, a bistable region between full defection and full cooperation appears already at $r=\ric(\theta, \delta_2, \delta_3)$, preceding the emergence of the interior branch $\rho^*_-$ at $r=\roc(\theta)$. The interior solution then terminates at \keepmath{r=\rsn(\theta,\delta_2,\delta_3)} inducing a discontinuous transition.
Qualitatively similar bistable regimes have recently been reported in single-order $\ell$-player nonlinear public goods games with heterogeneous synergistic and discounting interactions~\cite{zhu2024}. In that case, however, the group size is fixed and heterogeneity enters through the nonlinear character of the games, which may be synergistic or discounting with a given probability.
Here, by contrast, games of different number of players coexist, each with its own nonlinear response.
Thus, our results identify mixed-order games with competing nonlinearities as a distinct mechanism generating bistability between an active coexistence state and full cooperation.

Figure~\ref{fig:PD_mixed_l2_l3} shows the stationary density of cooperators in the $(\delta_3,r)$ plane for $\delta_2=0.5$ and different values of the fraction $\theta$ of three-player games. The limiting cases $\theta=0$ and $\theta=1$ recover the single-order two- and three-player games, respectively. For $\theta=0.1$, shown in Fig.~\ref{fig:PD_mixed_l2_l3}(b), the crossed-nonlinearity regime gives rise to a region where full cooperation and an active coexistence state are both stable (\mbox{C+ACT}). As $\theta$ increases, this bistable C+ACT region shrinks, as illustrated in Fig.~\ref{fig:PD_mixed_l2_l3}(c) for $\theta=0.25$, while the influence of triplet interactions becomes more pronounced and the phase diagram gradually approaches that of the single-order $\ell$-player nonlinear PGG. In addition, increasing $\delta_3$ drives the system from an active phase to full cooperation through a discontinuous transition, with the emergence of a bistable region between full defection and full cooperation.

To further illustrate the role of nonlinearities, Fig.~\ref{fig:PD_delta2_delta3} shows the stationary density of cooperators $\rho_{\mathrm{st}}$ in the $(\delta_2,\delta_3)$ plane for fixed $\theta$. This representation highlights the boundary between same-character and crossed-character regimes and identifies the parameter region where the saddle-node enters the physical domain. 

\begin{figure}[t]
    \centering    \includegraphics[width=0.49\textwidth]{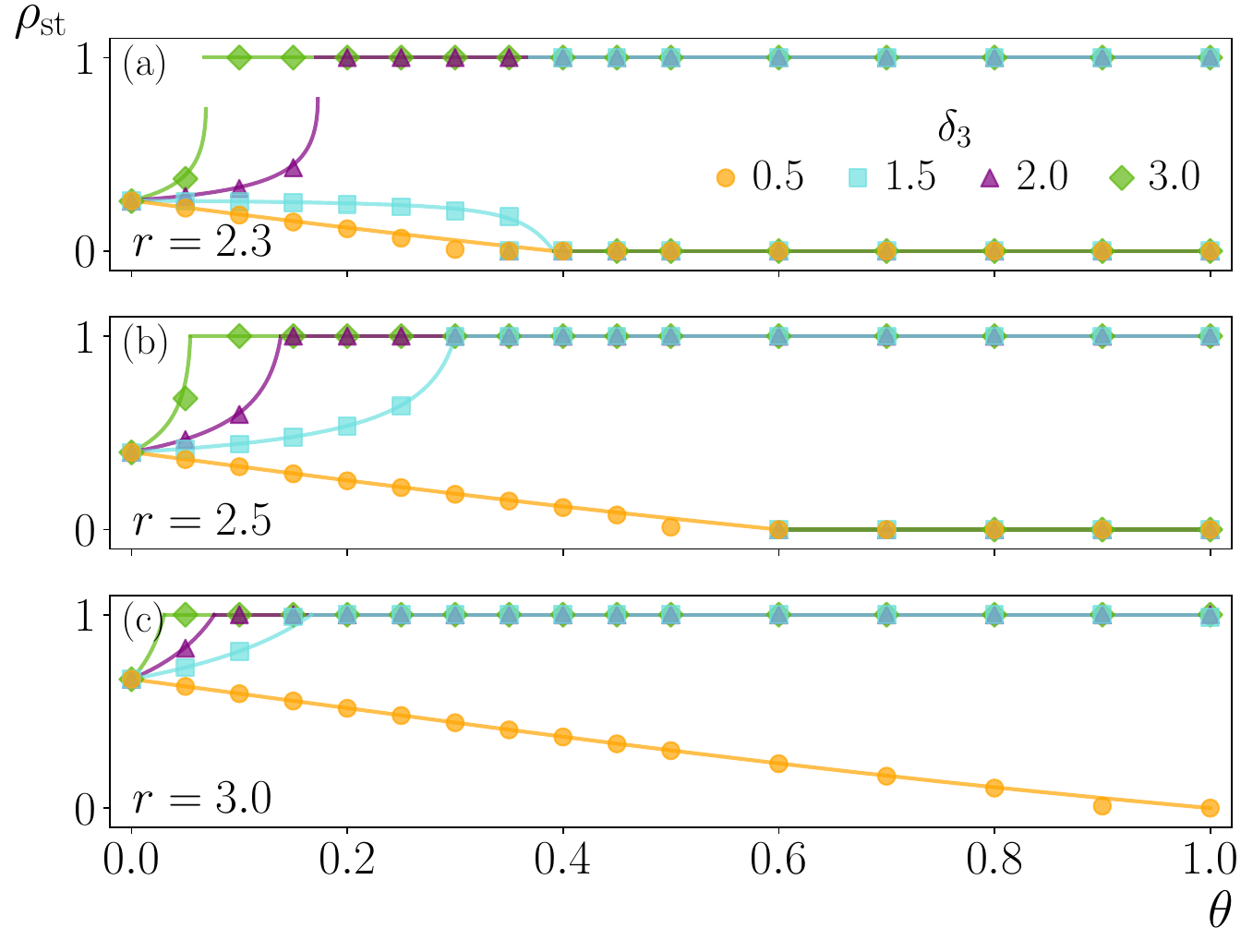}
    \caption{\textbf{Role of the fraction of three-player games in the stationary cooperation state.} Stationary density of cooperators $\rho_{\mathrm{st}}$ versus the fraction of three-player games $\theta$ for $\delta_2=0.5$ and several values of $r$ and $\delta_3$, as indicated in the legend. Symbols correspond to results obtained from computer simulations in the well-mixed population for $N=~12000$, while lines correspond to the analytical solutions of the mean-field replicator equation, Eq.~\eqref{eq:drhodt_mixed_23}.}
    \label{fig:rho_theta}
\end{figure}

Finally, Fig.~\ref{fig:rho_theta} summarizes the different transition scenarios obtained by varying the fraction of three-player games $\theta$. The figure shows the stationary density of cooperators $\rho_{\mathrm{st}}$ as a function of $\theta$ for three representative values of the benefit-to-cost ratio $r$ and several values of the triplet nonlinearity $\delta_3$. For $\theta=0$, all curves collapse, since in this limit the dynamics is purely pairwise and therefore independent of $\delta_3$. As $\theta$ increases, the contribution of triplet interactions becomes progressively more important. For a fixed value of $r>2$, the steady state of the system may evolve continuously towards full defection or full cooperation, or display a continuous increase followed by a discontinuous jump to the cooperative phase, depending on the value of $\delta_3$. We highlight that, for intermediate values of $r$, increasing $\theta$ can drive the system from the full cooperation phase to a bistable region consisting of full cooperation and full defection phases. This figure therefore illustrates how tuning the fraction of higher-order interactions not only changes the stationary cooperation level, but can also alter the nature of the transition.

These results show that the combination of interaction orders allows for a scenario in which nonlinearities compete with one another, which fundamentally alters the evolutionary dynamics and gives rise to dynamical regimes that are absent in single-order nonlinear PGGs.

\section{Structured populations} \label{sec:hypergraph}
In this section, we investigate how higher-order network structure affects the evolutionary dynamics of mixed-order two- and three-player nonlinear PGGs.

We consider hypergraphs composed of two- and three-hyperedges, corresponding to pairwise and triplet interactions, respectively, and analyze two qualitatively different classes of structured populations. First, we study random regular (RR) hypergraphs, in which every node participates in exactly $k_{/}$ and $k_\Delta$ two- and three-hyperedges, respectively. Second, we consider scale-free (SF) hypergraphs~\cite{lotito_hypergraphx_2023}, where the hyperdegree distributions of two- and three-hyperedges follow power laws with exponent $\gamma$. To ensure a fair comparison, we fix the average hyperdegree at each order for both types of hypergraphs. This allows us to disentangle the effect of structural connectivity patterns from those arising from the density of hyperedges.

Within the SF topology, we examine two structural features specific to heterogeneous systems. First, we analyze the influence of the spatial localization of the initial cooperative seed, comparing scenarios where cooperators are placed randomly, on nodes with the highest hyperdegree (hubs), or on nodes with the lowest hyperdegree (leaves). Second, we tune the inter-order hyperdegree correlation $\xi$ between two- and three-hyperedges. The case $\xi=1$ corresponds to maximal positive correlation, where nodes with the largest two-hyperdegree $k_{/}$ also have the largest three-hyperdegree $k_\Delta$. Conversely, $\xi=-1$ represents maximal anticorrelation, in which nodes that are highly connected at one order are minimally connected at the other.

\begin{figure*}[ht!]
    \centering
    \includegraphics[width=0.32\textwidth]{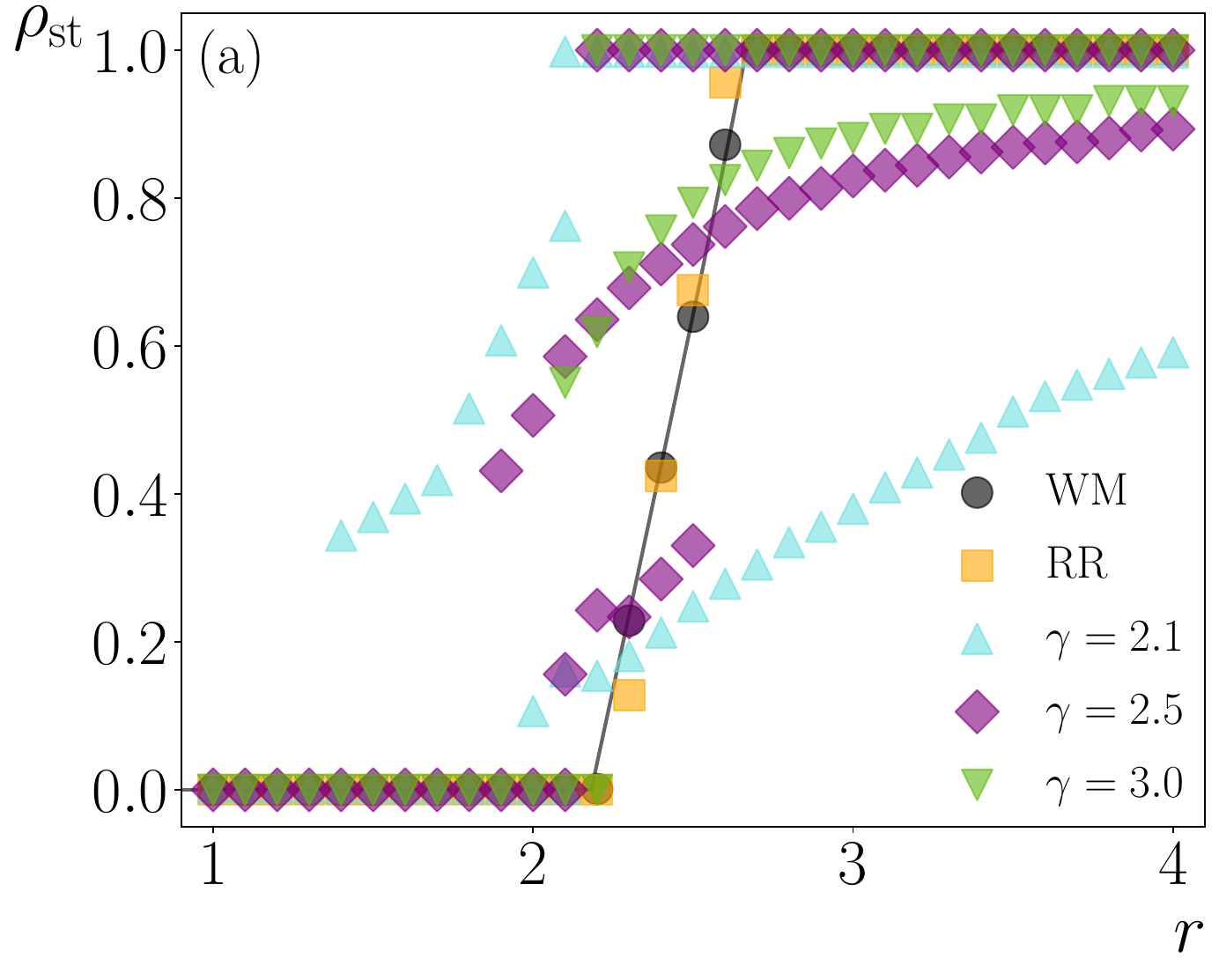}
    \includegraphics[width=0.32\textwidth]{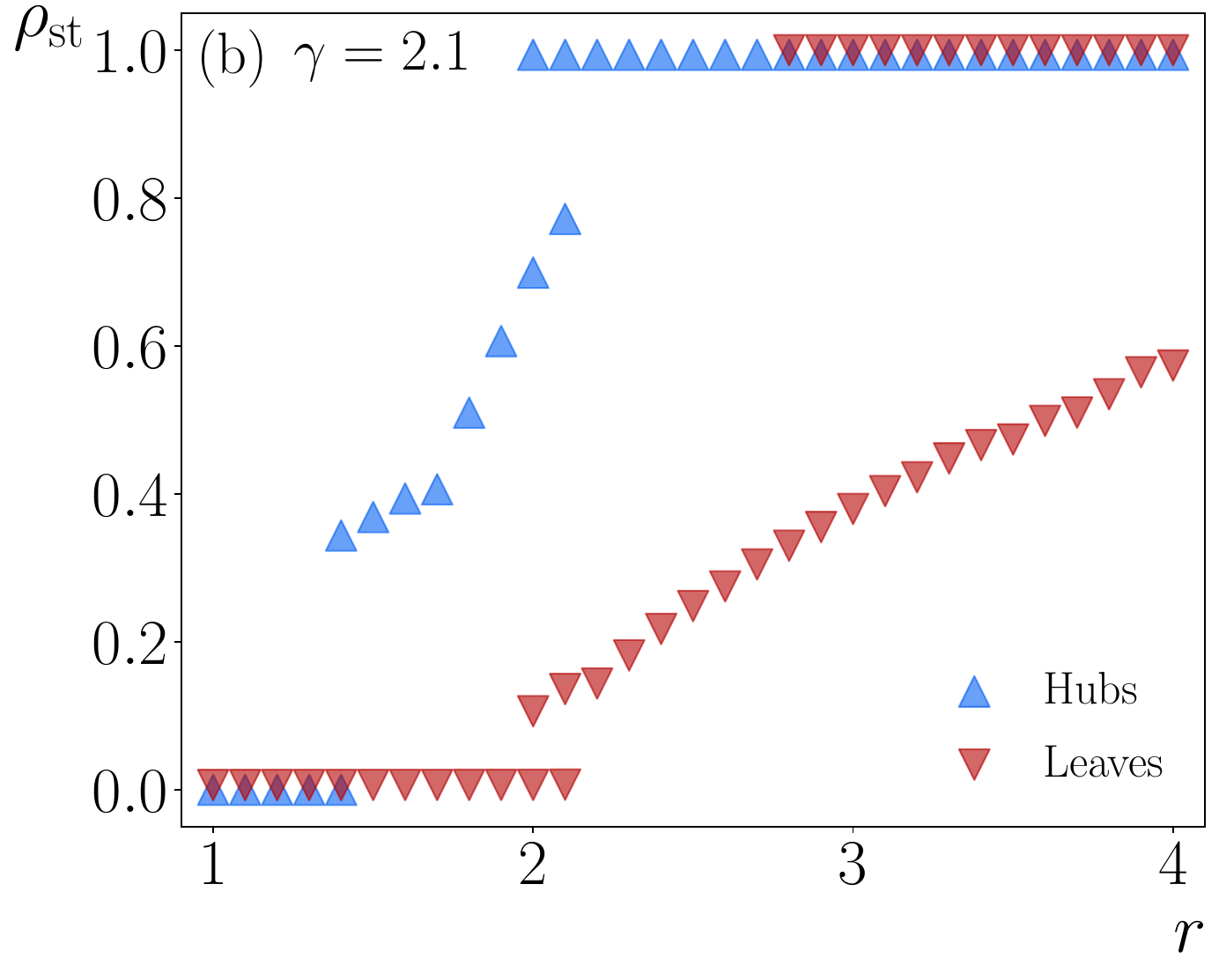}
    \includegraphics[width=0.32\textwidth]{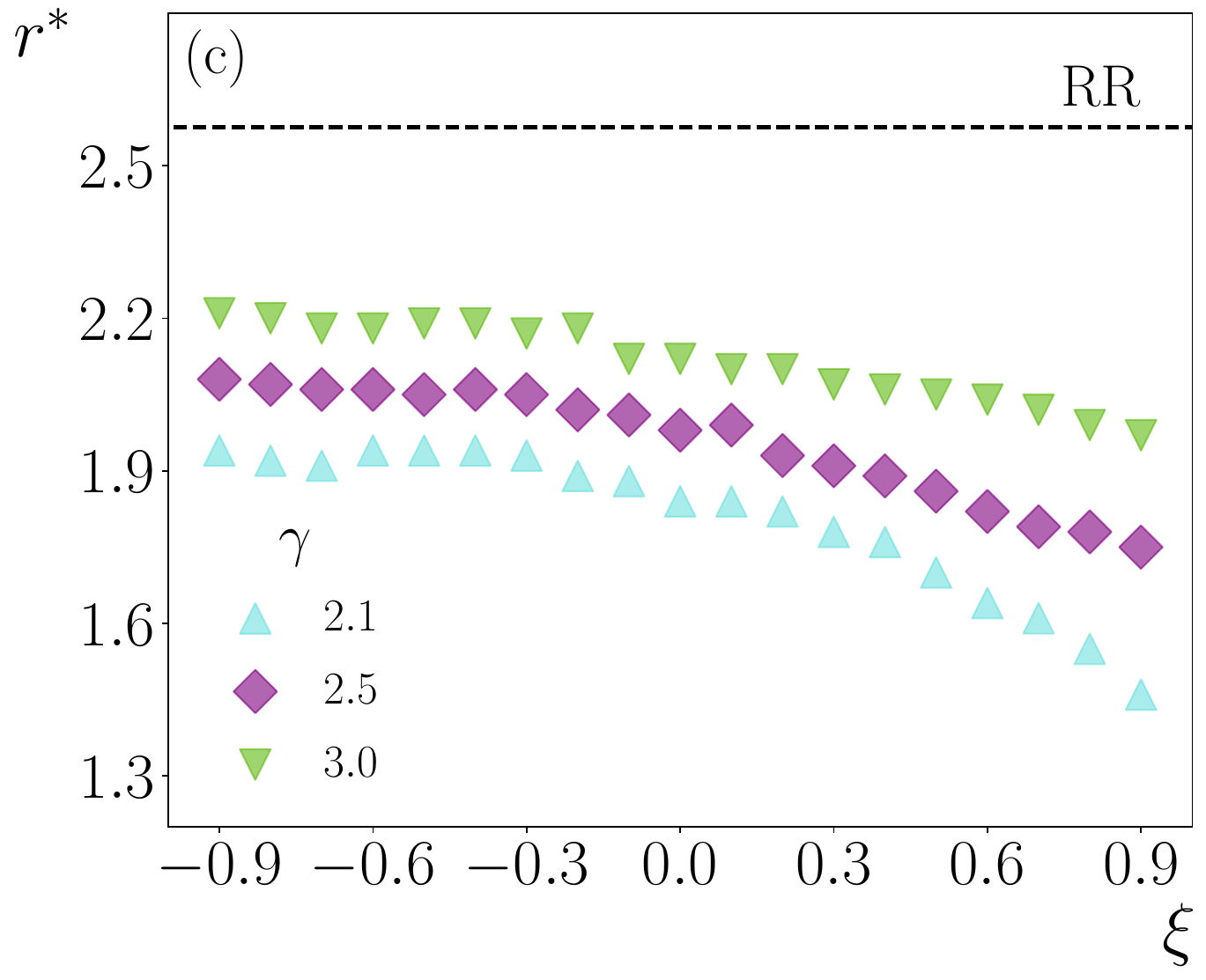}
    \caption{\textbf{Mixed-order two- and three-player nonlinear public goods games on hypergraphs.} (a) Stationary density of cooperators $\rho_{\mathrm{st}}$ as a function of the benefit-to-cost ratio $r$ for well-mixed (WM), random regular (RR), and scale-free (SF) hypergraphs with exponents $\gamma=2.1$, $2.5$, and $3.0$. The solid line represents the analytical solution of the mean-field replicator equation, Eq.~\eqref{eq:drhodt_mixed_23}. (b) Same observable for SF hypergraphs with $\gamma=2.1$, comparing different localizations of the initial cooperative seed $\rho(0)=0.05$ for highest hyperdegree (hubs), and lowest hyperdegree (leaves) nodes. (c) Critical benefit-to-cost ratio $r^\ast$ versus the inter-order hyperdegree correlation $\xi$ for an initial cooperative seed $\rho(0)=0.05$ placed at hubs on SF hypergraphs with exponents $\gamma=2.1$, $2.5$, and $3.0$. The critical ratio $r^\ast$ is defined as the smallest value of $r$ for which at least one stationary branch satisfies $\rho_{\mathrm{st}}>0.8$. The dashed horizontal line corresponds to the value $r^\ast$ for RR hypergraphs. Symbols correspond to results obtained from computer simulations on hypergraphs of size $N=2400$, with average two- and three-hyperdegrees $\langle k_{/}\rangle=6$ and $\langle k_{\Delta}\rangle=2$, and parameter values $\theta=0.25$, $\delta_2=0.5$, and $\delta_3=1.5$. In (a),(b) the interior branches are obtained from the quasistationary distribution (see Appendix~\ref{app:branch_extraction}).}
    \label{fig:structures}
\end{figure*}

Figure~\ref{fig:structures}(a) shows the stationary density of cooperators $\rho_{\mathrm{st}}$ as a function of the benefit-to-cost ratio $r$ for RR and SF hypergraphs, compared with the analytical results on well-mixed populations. 
For RR hypergraphs, the system closely follows the well-mixed predictions, indicating that homogeneous hypergraphs do not introduce significant deviations from the mean-field behavior in the present setting. This agreement should not be interpreted, however, as a generic equivalence between homogeneous structured populations and well-mixed populations, but rather as a consequence of using the pairwise comparison update rule.
The same phenomenon has been observed in previous studies for the prisoner's dilemma~\cite{alvarez2021evolutionary, civilini2024explosive}. In contrast, SF hypergraphs display qualitatively different behavior. First, cooperation is promoted, with a shift of the transition point towards smaller values of $r$. In addition, the nature of the transition itself is modified: besides the continuous transitions observed in well-mixed and RR hypergraphs, the system exhibits both continuous and discontinuous transitions with regions of bistability or even multistability, where two interior stationary states coexist with full defection or full cooperation. The location of these interior stationary states is identified from the quasistationary distribution, namely, the stationary distribution of $\rho$ conditioned on nonabsorption, as explained in Appendix~\ref{app:branch_extraction}. These effects arise from the heterogeneity of SF hypergraphs, where highly connected nodes can locally reinforce cooperative clusters and alter the global evolutionary dynamics. Our results highlight the critical differences between well-mixed and structured heterogeneous populations. In the mean-field limit, we previously assumed that the probability of finding $k$ cooperators in a group of $\ell$ players follows a binomial distribution (see Eq.~\eqref{eq:Binomial}) and that the probability of finding a cooperator is only dependent on the population level cooperators' density $\rho$. However, both of these assumptions can fail in the case of scale-free hypergraphs owing to the presence of hubs. In particular, hubs tend to cluster a particular strategy introducing correlations among players' strategies. Thus, the mean-field assumption of encountering a cooperator with a probability equal to population density $\rho$ breaks down. As a result, the insights that we have obtained so far for well-mixed populations do not hold any longer once we introduce scale-free hypergraphs and we get qualitatively different dynamical regimes.

Placement of initial strategies can drastically influence the stationary state of the evolutionary dynamics in either a positive or negative way. In particular, placing even a small fraction of cooperators on hubs can drive the system to higher levels of cooperation compared to the case in which they are placed randomly \cite{chen2008influence, kleineberg2017metric}. Even though previous studies have explored the role of placing cooperators on graphs \cite{poncela2007robustness, sadekar2024evolutionary}, similar effects are not studied on hypergraphs. Hypergraphs are particularly interesting because a hub can be defined using the nodes with the highest two- or three-hyperdegree or some combination of both. To illustrate the role of the localization of the initial cooperative seed in SF hypergraphs, Fig.~\ref{fig:structures}(b) shows the stationary density of cooperators $\rho_{\mathrm{st}}$ as a function of the benefit-to-cost ratio $r$ for SF hypergraphs with $\gamma=2.1$ when the initial fraction of cooperators $\rho(0)=0.05$ is placed either on hubs or on leaves. The localization of the cooperative seed strongly affects the emergence of cooperation. When cooperators are initially placed on hubs, cooperation spreads efficiently and the transition to full cooperation occurs at significantly smaller values of $r$. In contrast, initializing the system on leaves leads to a different behavior. In the latter case, the transition to full cooperation is noticeably delayed compared with the WM case. These results emphasize the role of structural heterogeneity in SF hypergraphs. In particular, highly connected nodes act as efficient spreading centers that facilitate the propagation of cooperative behavior.

Figure~\ref{fig:structures}(c) shows the threshold value $r^\ast$, defined as the minimum value of $r$ for which at least one stationary branch satisfies $\rho_{\mathrm{st}} > 0.8$, as a function of the inter-order hyperdegree correlation $\xi$ for an initial condition $\rho(0)=0.05$ placed at hubs on SF hypergraphs with different exponents $\gamma$. The dashed horizontal line indicates the value of $r^\ast$ for RR hypergraphs, which serves as a reference of homogeneity. We observe that $r^\ast$ decreases overall with $\xi$ for all values of $\gamma$, indicating that positive inter-order correlations systematically facilitate cooperation. For $\xi<0$, however, the threshold remains nearly constant, whereas for $\xi>0$ the decrease becomes more pronounced. This trend becomes stronger as the hypergraph becomes more heterogeneous. Moreover, for all values of $\xi$ considered, the threshold in SF hypergraphs remains below the RR benchmark, showing that structural heterogeneity promotes cooperation throughout the whole range of inter-order correlations.

\section{Discussion} \label{sec:discussion}
In this work, we studied nonlinear public goods games with mixed-order interactions in well-mixed and structured populations.

At the mean-field level, we showed that mixed-order interactions can generate a richer phase structure than single-order $\ell$-player nonlinear public goods games. More generally, for distinct interaction orders $\ell_1,\ell_2\geq2$, we found that at most two isolated interior fixed points can exist. Two qualitatively different situations emerge depending on the associated nonlinearities. When both interaction orders belong to the same regime, namely both discounting or both synergistic, the system reproduces the phenomenology of single-order $\ell$-player nonlinear public goods games, differing only in the location of the transition lines. By contrast, when the two interaction orders belong to opposite regimes, a saddle-node bifurcation may emerge and generate two interior fixed points, leading to a bistable regime that is absent in single-order games.

For the sake of simplicity and analytical tractability, we focused on the particular case $\ell_1=2$ and $\ell_2=3$, which allows for an explicit characterization of the stationary solutions and bifurcations. In this case, this mechanism gives rise to the coexistence between an active and a fully cooperative state, as well as to transition scenarios combining continuous and discontinuous features.
This comparison suggests that the emergence of the additional interior fixed points is closely linked to the competition between nonlinearities of opposite character. Our results show that mixed interaction orders provide a natural higher-order setting in which such competing nonlinearities can arise.
Qualitatively similar bistable regimes have recently been reported in single-order $\ell$-player nonlinear public goods games with heterogeneous synergistic and discounting nonlinearities~\cite{zhu2024}.
More generally, our results show that varying both the relative abundance of multiplayer games and their nonlinearity can modify not only the stationary level of cooperation, but also the nature of the transition itself.

We then showed that these effects persist, with important modifications, in structured populations. Random regular hypergraphs remain close to the mean-field prediction, indicating that homogeneous higher-order structures do not substantially alter the global phenomenology. Note that the close agreement between random regular hypergraphs and well-mixed populations might seem contradictory to the established results of games on graphs \cite{ohtsuki2006simple, nowak2006five}. However, owing to the differences in update rules and a qualitatively different mechanism for emergence of cooperation through mixture of games, this apparent contradiction is not true. Future studies could further explore the role of update rules in shaping the evolutionary dynamics. Heterogeneous hypergraphs, on the other hand, promote cooperation and shift the transition region in favor of cooperative behavior. In particular, scale-free hypergraphs enhance cooperation, broaden the range of parameters for which cooperative states can be sustained, and make the stationary state distributions dependent on the localization of the initial cooperators in the topology. Altogether, these results highlight that the structural organization of higher-order interactions shapes the collective dynamics of evolution of cooperation.

Overall, our work shows that nonlinear multiplayer games on hypergraphs already display a remarkably rich phenomenology once different interaction orders are allowed to coexist. The interplay between group size, nonlinearity, and structure provides a minimal but nontrivial setting in which new cooperative phases and transition scenarios emerge. In this sense, the present framework offers a natural starting point for studying more realistic evolutionary dynamics with multiple coexisting forms of social interaction.

\section*{Acknowledgments}
F.B. acknowledges support from the Italian Fund for Science (Fondo Italiano per la Scienza, FIS) of the Italian Ministry of University and Research (MUR), FIS 2 call, through the project CODYBE (FIS-2023-02989; CUP B53C25003310001), funded under Decreto Direttoriale No. 1236 of 1 August 2023 and admitted to funding by D.D. Prot. No. I.0018353 of 19 November 2025. F.B. and F.M. also acknowledge support from the Austrian Science Fund (FWF) through projects 10.55776/PAT1052824 and 10.55776/PAT1652425.
J.L. acknowledges the financial support received from Grants No. PID2021-122256NB-C21/C22 and No. PID2024-157493NB-C21/C22 funded by MICIU/AEI/10.13039/501100011033 and by ``ERDF/EU,'' and the María de Maeztu Program for units of Excellence in R\&D, Grant No. CEX2021-001164-M.

\bibliography{references.bib}

\appendix

\section{Interior stationary solutions in mixed-order $\ell_1$- and $\ell_2$-player nonlinear public goods games}
\label{app:mixed_order}

In this Appendix, we study the existence of interior stationary solutions of the mixed-order $\ell_1$- and $\ell_2$-player nonlinear public goods game in well-mixed populations. The dynamics is governed by the replicator equation
\begin{equation}
\frac{d\rho}{dt}=
\rho(1-\rho)F(\rho),
\end{equation}
where
\begin{align}
F(\rho)=&(1-\theta)\left(\frac{r}{\ell_1}
\left(
1+(\delta_{\ell_1}-1)\rho
\right)^{\ell_1-1}
-1\right)
\nonumber \\ &+
\theta \left(\frac{r}{\ell_2}
\left(
1+(\delta_{\ell_2}-1)\rho
\right)^{\ell_2-1}
-1\right),
\end{align}
such that the interior fixed points satisfy $F(\rho)=0$.

For determining the number of interior solutions, we introduce the auxiliary function
\begin{equation}
G(\rho)=
\frac{1-\theta}{\ell_1}
\left[
1+(\delta_{\ell_1}-1)\rho
\right]^{\ell_1-1}
+
\frac{\theta}{\ell_2}
\left[
1+(\delta_{\ell_2}-1)\rho
\right]^{\ell_2-1},
\end{equation}
such that $ F(\rho)=rG(\rho)-1$ and the interior fixed points therefore satisfy
\begin{equation}\label{eq:G}
G(\rho)=\frac{1}{r}.
\end{equation}

To determine the maximum number of interior fixed points, we therefore analyze the extrema of $G(\rho)$, which are given by the zeros of its derivative,
\begin{align}\label{eq:dGdrho}
\frac{dG}{d\rho}=&
\frac{(1-\theta)}
{\ell_1} (\ell_1-1)(\delta_{\ell_1}-1)
\left[
1+(\delta_{\ell_1}-1)\rho
\right]^{\ell_1-2}
\nonumber\\
&+
\frac{\theta}
{\ell_2} (\ell_2-1)(\delta_{\ell_2}-1)
\left[
1+(\delta_{\ell_2}-1)\rho
\right]^{\ell_2-2}.
\end{align}
If $\delta_{\ell_1}>1$ and $\delta_{\ell_2}>1$, both terms in Eq.~\eqref{eq:dGdrho} are strictly positive for all $\rho\in(0,1)$, implying that $G(\rho)$ is strictly increasing. Similarly, if $\delta_{\ell_1}<1$ and $\delta_{\ell_2}<1$, both terms are strictly negative and $G(\rho)$ is strictly decreasing. In both situations, $G(\rho)$ is monotonic and therefore the equation $G(\rho)=1/r$ can admit at most one interior solution as in the single-order $\ell$-player nonlinear public goods game.

The nontrivial situation arises when the two interaction orders belong to different nonlinearity regimes, namely \keepmath{\delta_{\ell_1}<1<\delta_{\ell_2}} or \keepmath{\delta_{\ell_2}<1<\delta_{\ell_1}}. In this crossed-nonlinearity case, the two contributions to $G'(\rho)$ have opposite signs, making the existence of interior extrema possible.
Without loss of generality, let us consider the case \keepmath{\delta_{\ell_1}<1<\delta_{\ell_2}}. An interior extremum of $G(\rho)$ must satisfy $dG/d\rho=0$, which gives the relation
\begin{equation}
\frac{
\left[
1+(\delta_{\ell_2}-1)\rho
\right]^{\ell_2-2}
}{
\left[
1-(1-\delta_{\ell_1})\rho
\right]^{\ell_1-2}
}
=
\frac{(1-\theta)}{\theta}
\frac{
\ell_2(\ell_1-1)
}{
\ell_1(\ell_2-1)
} \frac{(1-\delta_{\ell_1})}{(\delta_{\ell_2}-1)}.
\label{eq:extremum_condition_general}
\end{equation}
Note that the right-hand side of Eq.~\eqref{eq:extremum_condition_general} is independent of $\rho$. Hence, the number of interior extrema is determined by the behavior of the left-hand side. The numerator is nondecreasing and the denominator is nonincreasing throughout the physical interval $\rho\in(0,1)$. Since the interaction orders are distinct and at least two, at least one of these dependences is strictly monotonic. Consequently, their ratio is strictly increasing and Eq.~\eqref{eq:extremum_condition_general} can be satisfied at most once. It follows that $dG/d\rho$ possesses at most one zero in the physical interval and, therefore, $G(\rho)$ can have at most one interior extremum. Since the interior stationary solutions are given by the intersections between $G(\rho)$ and the horizontal line $1/r$, we conclude that the mixed-order nonlinear public goods game can admit at most two interior stationary solutions in the physical interval $\rho\in(0,1)$.

\section{Details of the stochastic simulations. Extraction of stationary branches from temporal trajectories}
\label{app:branch_extraction}
The stochastic evolutionary dynamics on finite-sized scale-free hypergraphs may converge not only to the absorbing states at $\rho^*=0$ and $\rho^*=1$, but also to other interior stable branches. Finite systems showcase fluctuations around these interior solutions until they reach one of the two absorbing states. To identify these interior states correctly, we study the \textit{quasistationary} distribution, formally defined as 
\begin{equation}
    P_\mathrm{qst}(\rho)=\lim_{t\to\infty}P(\rho,t|\rho\neq0, \rho\neq1),
\end{equation}
which captures the long-term behavior of a system when it has not reached either absorbing state.

For each set of structural parameters $(\theta,\xi,\gamma)$, we generate $N_{\mathrm{HG}}=10$ independent hypergraphs. For each set of dynamical parameters $(r,\delta_2,\delta_3)$ and each of the $10$ initial conditions uniformly distributed in the interval $\rho(0)\in[0.05,0.95]$, we analyze $n_{\mathrm{traj}}=100$ independent temporal trajectories $\rho(t)$ for each hypergraph. For every trajectory $\rho(t)$, we retain only the steady state $t\in[T,t_{\mathrm f}]$ of the time series, after discarding
the initial transient state. If the system has reached one of the two absorbing states, the trajectory is accordingly assigned to the lower branch (full defection) or to the upper branch (full cooperation). Otherwise, it is classified as an active trajectory. 

To identify the interior stationary branches, we first merge all the steady state of the trajectories classified as active. From these samples, we construct the probability distribution of $\rho$, which is precisely the quasistationary distribution $P_{\mathrm{qst}}(\rho)$ since it has been built from non-absorbing trajectories. For each detected interior peak, we determine the interval around the peak in which the quasistationary distribution remains above $60\%$ of the peak height. If this width is smaller than $0.1$, we use the peak position as the branch value. Otherwise, we use the weighted median of the interval. This provides a more robust representative value of the interior points in cases where the quasistationary distribution displays a broad plateau rather than a sharp maximum.

The absorbing branches are retained only when their empirical weights exceed a minimum threshold. Let $p_0$ and $p_1$ denote the fractions of trajectories assigned to the lower and upper absorbing states, respectively. These branches are included in the stationary diagram only when $p_0$ or $p_1$ is larger than a threshold $p_\mathrm{c}=0.1$.

For each set of structural and dynamical parameters mentioned above, a collection of stationary branch positions $\{\rho^\ast\}$ is obtained. It may contain the two absorbing solutions as well as one or two interior stationary branches.

\end{document}